\documentclass{elsarticle}
\usepackage{amsmath}
\usepackage{amssymb}
\usepackage{graphicx}
\usepackage{dcolumn}
\usepackage{bm}

\begin{document}
\title{Topological end states and Zak phase of rectangular armchair ribbon}

\author{Y. H. Jeong}

\author{S. -R. Eric Yang\corref{cor1}}
\cortext[cor1]{Corresponding author}
\ead{eyang812@gmail.com}

\address{  Department of Physics, Korea  University, Seoul,
Korea\\}

\begin{abstract}
We consider the end  states of a half-filled rectangular armchair graphene ribbon (RAGR) in a staggered potential.
Taking electron-electron interactions into account we find that,  as  the strength of the staggered potential varies,
three    types of couplings between the end states can occur:   antiferromagnetic  without or with spin splitting,  and paramagnetic  without spin-splitting.   We find that a spin-splitting is present only in the  staggered potential region $0<\Delta<\Delta_c$.  The transition from the antiferromagnetic state at $\Delta=0$    to the paramagnetic state goes through an intermediate spin-split antiferromagnetic state, and this  spin-splitting disappears suddenly at $\Delta_c$.
For small and large values of $\Delta$  the end charge of a RAGR can be connected to the Zak phase of the periodic armchair graphene ribbon (PARG) with the same width, and it varies continuously as  the strength of the potential changes.

\end{abstract}

\maketitle

\section{Introduction}

Graphene  exhibits interesting fundamental physics\cite{Neto}, such as  quantum Hall effect\cite{PKim}, Berry phases\cite{Ando}, and edge magnetism\cite{Fuj,Son}.
End states, located at the end points of a long  insulating one-dimensional wire\cite{Kit}, can reflect the presence of   gap states with fractional properties and the nature of various electronic and magnetic bulk states.   These objects are found  in polyacetylene, spin chains, Kondo insulator, and other systems\cite{Sol,Guo,Ng,Piers,Jeong}.  One may also ask whether an end charge exists in quasi-one-dimensional graphene ribbons  and how it may be  related to the Zak phase of the underlying band structure\cite{Zak,Van1}.

\begin{figure}[!hbpt]
\begin{center}
\includegraphics[width=0.7\textwidth]{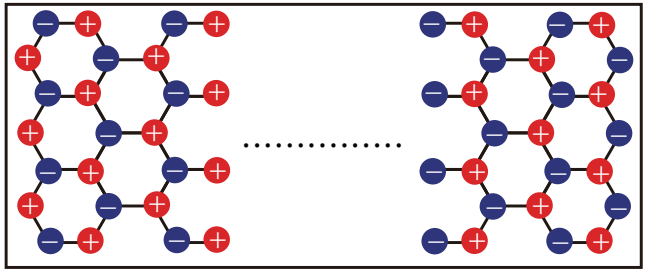}
\caption{ RAGR has  two long armchair edges and two short zigzag edges.  The lengths of  the armchair and zigzag edges are, respectively, $L_{arm}$ and $L_{zig}$ (ribbon width $W$ is equal to $L_{zig}$).   The staggered potential energy is  $\epsilon_i=\Delta/2$ on sublattice A (red circles)  and $\epsilon_i=-\Delta/2$ on sublattice B (blue circles).  Left (right) end is made of A (B)-type sites.}
\label{stag}
\end{center}
\end{figure}

A PAGR with the width $ (3L + 1)a_0$ or $3La_0$
has an energy  band gap and is semiconducting\cite{Brey,Yang,Loss,exptarm,Lee}  ($L$ is an integer and $a_0$ is the graphene unit cell length).  On the other hand, an armchair ribbon with the width $(3L + 2)a_0$ and a zigzag ribbon do not have a gap in the absence of electron-electron interactions\cite{Yang}  (we will not consider these ribbons here).
It is useful to think about a RAGR  as generated from such a   PAGR by cutting it transversely.
A  RAGR has two long armchair edges and two short zigzag edges\cite{Tang,Kim}, see Fig.\ref{stag}.
The left end of the RAGR is made of the  A-type zigzag edge while the  right end  is made of the B-type.
Gap states  appear  that are localized on the end sites\cite{Jeong}, and their  number   grows with increasing  length of the zigzag edges. In our work we will refer
to  these gap states as end states.
Since the magnitude of the energy gap is sizable these states  are isolated from other lower and higher energy quasi-continuum  states.    A staggered potential\cite{stagpot1,stagpot2,stagpot3,Sor} may modify   the properties of end states of a RAGR, and   the relation between the Zak phase of the PAGR and the end charge may also change.

Our RAGR is assumed to be half-filled with an energy gap and its ribbon width is such that four spin-resolved gap states are present.  Two of them are occupied while  the other  two  are unoccupied.   We find that the properties  of the end states depend on the interplay between the  strength of staggered potential, on-site repulsion, and  ribbon length and  width.
As the strength of the staggered potential changes,   we find  three coupling types between the end states:   antiferromagnetic  without or with spin-splitting,  and paramagnetic without spin-splitting.   In this paper  antiferromagnetic (ferromagnetic) gap  states  mean that  their spins on opposite zigzag edges are coupled antiferromagnetically (ferromagnetically).
Paramagnetic means zero net spin.
The spin-splitting   $\delta$ is   defined as the energy difference between
the highest occupied spin-up and -down gap states.
Spin-splitting of the gap states is present only in the staggered potential region $0<\Delta<\Delta_c$ (quasi-continuum states near  the energy gap are also  spin-split by  a small amount).  The transition from the antiferromagnetic state at $\Delta=0$    to the paramagnetic state at $\Delta_c$ goes through an intermediate antiferromagnetic state whose gap states are spin-split.  This spin-splitting vanishes abruptly at $\Delta_c$.  For $\Delta/t\gg1$, where the hopping parameter is $t\sim 3$eV,  and for some small $\Delta$      the value of the end charge  can be connected to the Zak phase of the PAGR  with the same width, and  the end charge varies continuously as the strength of the potential changes.

\section{Band structure and Zak phase  in a staggered potential  }

We would like to connect the end occupation numbers of a RAGR  to  the Zak phase of the band structure of the PAGR with the same ribbon length.  We will investigate under what conditions they can be connected.
We adopt a tight-binding model with  on-site repulsion $U$ and compute the band structure  using the Hartree-Fock approximation(HFA)\cite{Stauber}.   This approach is widely used and its results are consistent with those of DFT\cite{Fuj,Yang,Pis}.

\begin{figure}[!hbpt]
\begin{center}
\includegraphics[width=0.7\textwidth]{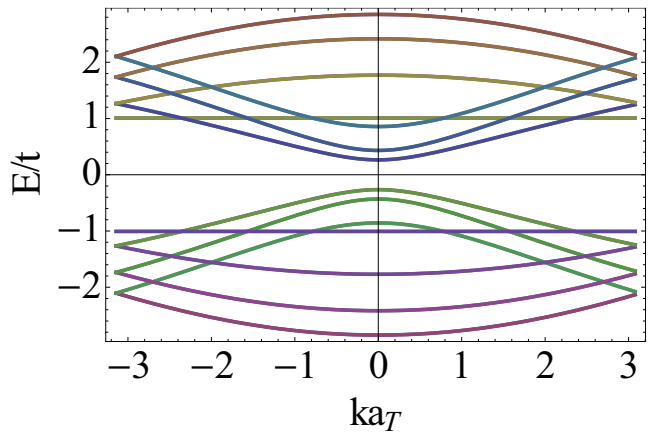}
\caption{ Band structure of a half-filled  PAGR  with width  $W=3a_0$ and the  length of armchair  edges    $L_{arm}=215.84$ \AA) (Here $a_0$ is the unit cell length of the honeycomb lattice.  The Fermi energy is $E_F=0$).  On-site potential is $U=0.5t$ and staggered potential is $\Delta=0.3t$. The length of the unit cell of the periodic ribbon is $a_T=3a_{cc}$, where $a_{cc}$ is the carbon-carbon distance. }
\label{stagband}
\end{center}
\end{figure}

Figure \ref{stagband} displays the HF band structure $E(k)$ in the presence of a staggered potential.   The ground state is paramagnetic for various values of $\Delta$ and $U$ used in this work,  and no spin splitting of the occupied bands  and no magnetization on the armchair edges are found.  Bulk graphene band structure also does not show a band spin-splitting in a staggered potential\cite{stagpot1}.  But a periodic zigzag graphene ribbon does display a spin-splitting\cite{Sor}, and the presence of zigzag edges is important for the spin-splitting (note  that our PAGR has no   zigzag edges and end points).
The size of the gap  increases as  the strength of the staggered potential increases.   The computed band structure $E(k)$ is for the wave vector $k$ is parallel to the ribbon axis.   Note that the band structure has  band crossings.

" The band crossings and numerically computed eigenstates can give to   states with  wildly different phases, making it difficult to compute numerically the Zak phase.
Here we use a gauge invariant method\cite{Resta}.
Using the periodic part of the Bloch wavefunctions  of the occupied  bands we compute the    spin-independent Zak phase
\begin{eqnarray}
\label{eq:zakphse}
\mathcal{Z}/2\pi
&=&\frac{1}{2\pi}\sum_{s=0}^{N_s-1} \mathrm{Arg}\left[   \mathrm{det}  \langle C_{l,k_s}| C_{l',k_{s+1}} \rangle   \right].
\label{Zakphase}
\end{eqnarray}
The expansion coefficients $C_{l,k}$ are column vectors, whose components are the site indices  $i=(m,A)$ or $(m,B)$, labeling  atoms in the unit cell.  They are obtained from our  tight-binding Hamiltonian
matrix,.  Here, the Brillouin zone is divided into small intervals labeled by $k_s$.  For each $k_s$, one computes all the  matrix elements $\langle C_{l,k_s}| C_{l',k_{s+1}} \rangle $ (the matrix index $l$ runs over the occupied bands).
\begin{figure}[!hbpt]
\begin{center}
\includegraphics[width=0.7\textwidth]{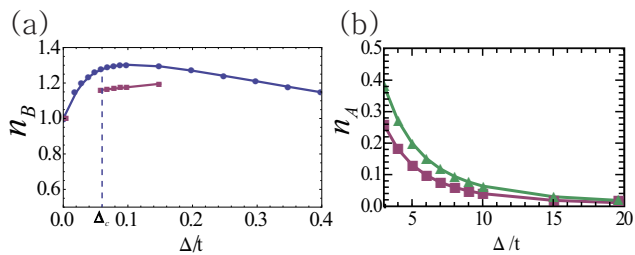}
\caption{ (a)  Charge per right end site  computed from  the Zak phase  (circles) and  from the occupation number  (squares) in the region    $\Delta/t<1$.   (b)  Charge per left end site   computed from the Zak phase  (triangles)  and from the occupation number (squares)  in the region   $\Delta/t\gg1$.     Parameters are identical to those of Fig.\ref{stagband}.   }
\label{Zak}
\end{center}
\end{figure}

For an  {\it insulating interface} of a RAGR the  total end charge  per spin is related to the Zak phase of the PAGR:
$Q/e=\frac{Z}{2\pi}$\cite{Van1}.   Note that  due to gauge invariance the  phase $Z$ is only  defined within mod $2\pi$.
The occupation number per site $n_{A,B}$ is obtained by dividing $Q/e$ with  the total number $N$ of  A or B end sites.      The condition for an insulating interface
is that   the average end occupation number  per A-site  is  $n_{A}\approx 0$ or $n_{A}\lesssim1$, corresponding, respectively, to $\Delta/t\gg1$ or $\Delta\rightarrow 0$ (the condition for the B-sites is $n_{B}\gtrsim1$ or $n_{B}\approx 0$).
Figure \ref{Zak} shows results of  $n_B$ computed from the Zak phase  and from the numerical calculation of the occupied wavefunctions.     As shown in Fig.\ref{Zak} (a),
at  $\Delta=0$ they agree with each other: $n_B=1$ (the Zak phase of each spin is $0$ mod $2\pi$).
In the region   $\Delta/t\gg1$ the two agree  as  $\Delta/t$ increases, see Fig.\ref{Zak} (b).
However, for $0<\Delta<\Delta_c$ our numerical result indicates that
the comparison between the two approaches cannot be made ($\Delta_c=0.06t$).   This is because, the gap states  and the   quasi-continuum states of the RAGR near the gap  are   spin-split  while  all the states of the PAGR are spin degenerate (spin-splitting and occupation numbers are  computed below). In the region   $\Delta \gtrsim\Delta_c$ the  agreement between the two approaches for the B-end site is within  $15-20\%$, as shown in Fig.\ref{Zak} (a).   Note that the value of the occupation number is somewhat larger than one: the spin down ($\sigma=-1$) part  contributes $1$ to $n_B$ and the opposite spin ($\sigma=1$) part  accounts for the remaining part.

\section{ Antiferromagnetic and paramagnetic couplings}

In RAGRs  the antiferromagnetic, paramagnetic, antiferromagnetic with broken sublattice symmetry,
and ferromagnetic states compete against each other.
To compute the end charges, magnetization, and spin-splitting of our RAGR we use the following approach.
Since translational symmetry is broken we write  a tight-binding Hamiltonian  in the site representation  including  the on-site repulsion $U$.  We solve it using the Hartree-Fock (HF)  approximation (the dimension of the Hamiltonian matrix is  $\sim1400$). In general the nature of the ground state depends on the interplay between several parameters $\Delta$, $U$, $t$, $L_{zig}$,  and $L_{arm}$. We investigate  RAGRs with a short width, $\frac{L_{zig}}{L_{arm}}\ll1$.  The end electrons interact with the bulk electrons and a many-body calculation is required.   We compute the average end occupation number  per A-site of   the left zigzag edge:
$ n_A=\sum_{\sigma}n_{A,\sigma}=\frac{1}{N}\sum_{\sigma}\sum_{\alpha\in occ,i}|\psi^{\alpha}_{A,\sigma}|^2$, where the sum is  over   the occupied states $\alpha$, spin states $\sigma$,  and site index $i$ ($\psi^{\alpha}_{i,A}$ are the  probability amplitudes).  Note that, in addition to occupied gap states, there are also occupied quasi-continuum states that have to be included in the sum.  At a B-site on the opposite end  the occupation number  $ n_{B}$   is defined similarly.    The average per site magnetizations  on the left and right zigzag edges are, respectively,   $m_A=n_{A,\uparrow}-n_{A,\downarrow}$ and $m_B=n_{B,\uparrow}-n_{B,\downarrow}$.

\begin{figure}[!hbpt]
\begin{center}
\includegraphics[width=0.7\textwidth]{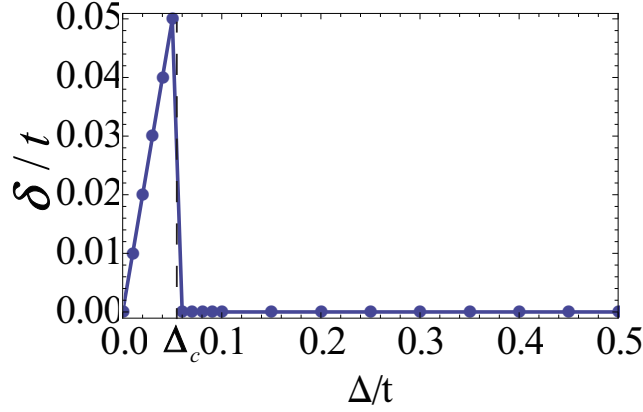}
\caption{Spin-splitting of gap states $\delta$ of  a RGAR is  shown ($\delta$ is defined in Sec.1 ).  On-site potential is $U=0.5t$,     width $L_{zig}=3a_0$, and  ribbon length   $L_{arm}=215.84$ \AA.}
\label{spinsp}
\end{center}
\end{figure}

In Fig.\ref{spinsp} we have computed, as a function of $\Delta$, the spin-splitting of the gap states of the  ground state. It indicates  that a sudden  electronic and magnetic  reconstruction   of the end states occurs near $\Delta_c$.
At $\Delta=0$ we find that the ground state is antiferromagnetic without spin-splitting (however,  there is a ferromagnetic state whose energy nearly degenerate\cite{Tang}).  In the interval $0<\Delta<\Delta_c$ the ground state is antiferromagnetic  but with spin-splitting.     For
$\Delta>\Delta_c$  the ground state is paramagnetic without spin-splitting.

\begin{figure}[!hbpt]
\begin{center}
\includegraphics[width=0.7\textwidth]{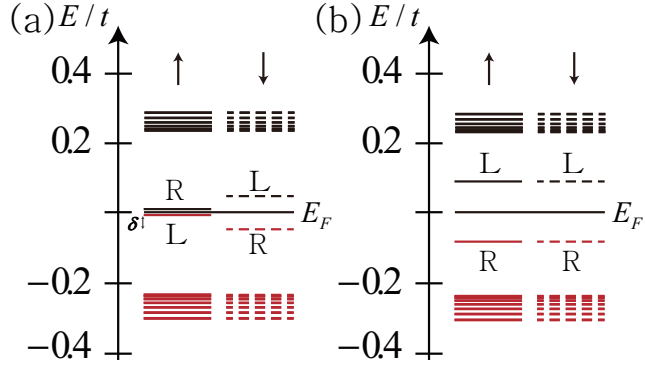}
\caption{ Schematic drawing of energy spectrum for spin-up and -down end/gap  states of a  RAGR with width $L_{zig}=3a_0$.    Two values of the strength of the staggered potential,
$0.03t$ (a)   and $0.3t$ (b), are used.   The first value  is just below the critical value  $\Delta_c$  and the second  one is far above it.
Spin splitting $\delta$ is present  in (a)   while not in (b). In (a) the occupied  end  states are located on the opposite zigzag edges (see Fig.\ref{endch}(a)).   While in (b) they are both located on the right zigzag edge (see Fig.\ref{endch}(b)).  Symbols $R$ and $L$ mean that    an end   state is localized, respectively, on the right and left zigzag edges. }
 \label{ES}
\end{center}
\end{figure}

\begin{figure}[!hbpt]
\begin{center}\includegraphics[width=0.7\textwidth]{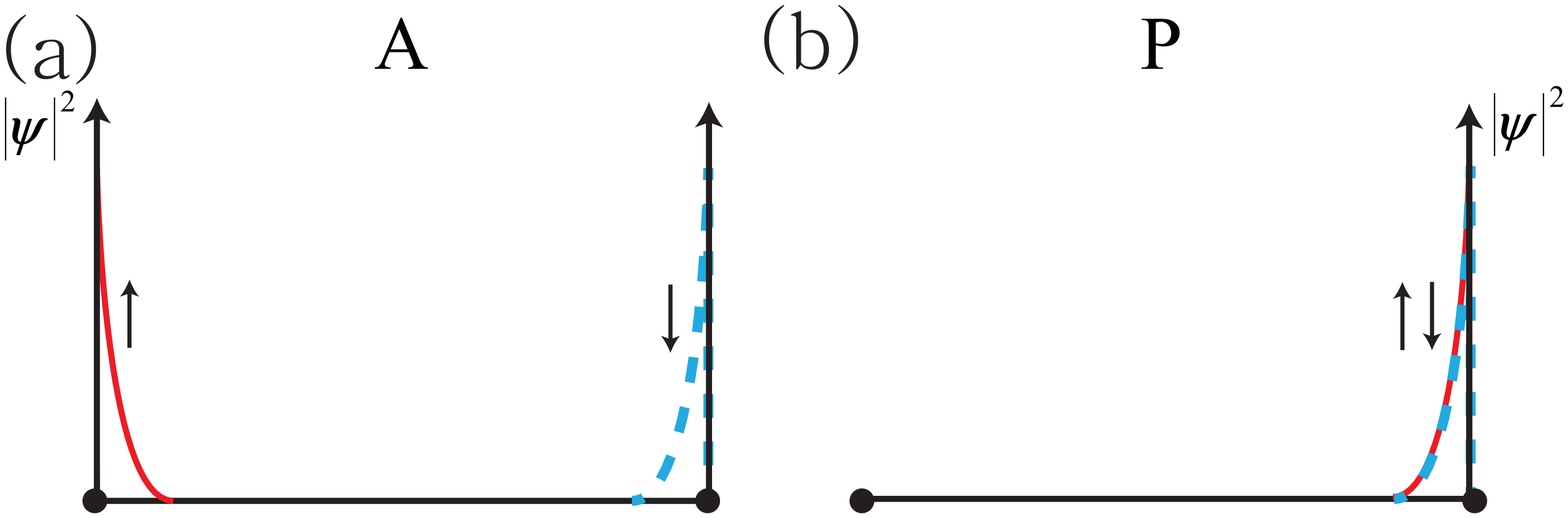}
\caption{Schematic probability densities of the occupied   end  states are displayed. They are located either on the left or right end sites. Staggered potential energy is lower on the right end sites.  (a) End  states are located on opposite end sites  with antiparallel  spins: antiferromagnetism;
(b) end  states are located on the same  end sites  with antiparallel  spins: paramagnetism. }
\label{endch}
\end{center}
\end{figure}

To explain the abrupt change   we need to investigate what types of gap states are occupied as $\Delta$ changes.
Figure \ref{ES} displays the computed energy spectra for spin-up and -down gap states and quasi-continuum states outside the gap.   Figure \ref{endch}   displays the schematic probability densities of the  occupied gap states.
For $\Delta<\Delta_c$  the  spin-up and -down gap states are  located, respectively, on the left  and right   end sites  (see Figs.\ref{ES}(a) and \ref{endch}(a)).    In this case the occupied spins of the
gap states are coupled antiferromagnetically.   Note that near the critical
value $\Delta_c$ the spin-up gap states are nearly degenerate, see Fig.\ref{ES}(a).   On the other hand, for $\Delta>\Delta_c$ both are localized on the right end sites, as shown in Figs.\ref{ES}(b) and \ref{endch}(b).  The antiferromagnetism vanishes.
At the critical value $\Delta_c$ an electron  localized on the left edge,  transfers to the right end sites, where    the potential energy is lower.   However, there is also an accompanying rearrangement of the occupied quasi-continuum states, and the resulting total charge transfer to the right end sites may not be entirely  one.    For  $\Delta\gtrsim \Delta_c$      the occupation per site is somewhat less than $n_B=4/3$, but it quickly approaches this value as $\Delta$ increases and the total
electron number  on the three  B-end sites  becomes  $4$, consistent with  one extra electron on the right edge.
The  magnetization  $m_A$ is almost not changing until   $\Delta_c$, where   it   suddenly  decreases to zero from a positive value.  Also the  magnetization
$m_B$ is not changing until   $\Delta_c$, where   it    increases  abruptly to zero from a negative value.  Below $\Delta<\Delta_c$   the sum is $m_A+m_B=0$, consistent with antiferromagnetism.


\section{Summary and discussions}

We have shown that for  the values of the  staggered potential  $\Delta/t\gg1$ and for some small $\Delta$ the end charge of a RAGR  can be connected to the Zak phase of the PAGR with the same width.  The computed end charge varies continuously on the strength of the potential.   The comparison between the two quantities  cannot be made in the region $0<\Delta<\Delta_c$  because the gap states and the quasi-continuum states near the gap  of the RAGR display  spin-splitting  while  those of the   PAGR  do not.

As the strength of the staggered potential varies  we find that the end states of a RAGR can be divided into three magnetic coupling regimes:   antiferromagnetic  without or with spin-splitting,  and paramagnetic without spin-splitting.  The transition from the antiferromagnetic state at $\Delta=0$    to the paramagnetic state goes through an intermediate antiferromagnetic state with  spin-splitting. The spin-splitting disappears suddenly at $\Delta_c$.

We have performed the HF calculations with larger widths and lengths of the rectangular ribbon than those used in Fig.\ref{Zak}.  We find that, although a  longer zigzag edge produces more gap states, the dependence of the end charge on the strength of the staggered potential is qualitatively similar to the   result for shorter zigzag edges:  at $\Delta=0$ the occupation number per site is one and at a critical value $\Delta_c$ paramagnetism appears.  In the absence of the staggered potential ($\Delta=0$) inversion symmetry is intact and  the Zak phase is $0$ mod $2\pi$, independent of the range of electron-electron interactions.   At $\Delta\neq 0$, where inversion symmetry of the electronic density of the ground state  is broken, there may be some quantitative differences between the results of the short-range and long-range interactions.  But, as we  mentioned before, the results of a short-range model are expected to be  in qualitative agreement with those of a first principles calculation of graphene nanoribbons  using the long-range Coulomb interactions, see refs. \cite{Yang,Pis}.   Moreover, the HF treatment of the long-range Coulomb interactions in bulk two-dimensional graphene leads to the renormalization of the band structure with only slight deviations from the linear behavior with unchanged electronic wavefunctions\cite{Stauber}.

It may be  worthwhile to  find ways to modulate the end/gap    states by strain or electric field.      The measurement of the differential conductance, using scanning tunneling microscopy\cite{Andrei}, may  provide rich information on  the end states.

\section*{Acknowledgments}
This research was supported by Basic Science Research Program
through the National Research Foundation of Korea(NRF) funded by the
Ministry of Education, ICT $\&$ Future Planning(MSIP) (NRF-2015R1D1A1A01056809).

\end{document}